\documentclass[%
reprint,
aps, twocolumn,
superscriptaddress,
bibnotes,
 amsmath,amssymb,
 aps,
numerical
]{revtex4-2}
\usepackage[utf8]{inputenc}
\usepackage{comment}
\usepackage{natbib}
\usepackage{todonotes}
\usepackage{float}
\usepackage{etoolbox}
\usepackage[utf8]{inputenc}
\usepackage{siunitx}
\usepackage{array}
\usepackage{physics}
\usepackage{dsfont}
\usepackage[english]{babel}			
\usepackage{xcolor}
\usepackage{graphicx}				
\usepackage{amssymb,amsmath}		
\usepackage{url}					
\usepackage{marvosym}				
\usepackage[T1]{fontenc}            %
\usepackage{wrapfig}				
\usepackage{charter} 				
\usepackage{blindtext}				
\usepackage{datetime}				
\usepackage{lipsum}                 
\usepackage{float}                  
\usepackage{tabularx}
\usepackage{siunitx}
\usepackage{dcolumn}
\usepackage{bm}
\usepackage{todonotes}
\usepackage{hyperref}
\hypersetup{colorlinks=true,allcolors=blue}

\newcommand{\UIBK}{Institut f{\"u}r Experimentalphysik, Universit{\"a}t Innsbruck, 6020 Innsbruck, Austria}

\newcommand{\TUdo}{Condensed Matter Theory, Department of Physics, TU Dortmund, 44221 Dortmund, Germany}

\newcommand{\JKU}{Institute of Semiconductor and Solid State Physics, Johannes Kepler University Linz, Linz, Austria}

\newcommand{\ACP}{Institute of Applied Physics, Abbe Center of Photonics, Friedrich Schiller University Jena, 07745 Jena, Germany}
\newcommand{\IOF}{Fraunhofer Institute for Applied Optics and Precision Engineering IOF, Center of Excellence in Photonics, 07745 Jena, Germany}
\newcommand{\TuB}{Institute of Solid State Physics, Technische Universität Berlin, 10623 Berlin, Germany}

\newcommand{\Cam}{Current address: Cavendish Laboratory, JJ Thomson Avenue, University of Cambridge, CB3 0HE Cambridge, UK}

\newcommand{\brazil}{Universidade Estadual de Campinas, Instituto de F\'{\i}sica Gleb Wataghin, 13083-859 Campinas, Brazil}

\preprint{APS/123-QED}

\begin{document}

\title{Robust Single-Photon Generation for Quantum Information Enabled by Stimulated Adiabatic Rapid Passage}

\author{Yusuf Karli}
\thanks{These authors contributed equally}
\affiliation{\UIBK}
\affiliation{\Cam}
\author{Ren\'e Schwarz}
 \thanks{These authors contributed equally}
\affiliation{\UIBK}
\author{Florian Kappe}
\thanks{These authors contributed equally}
\affiliation{\UIBK}
\author{Daniel A. Vajner}
\affiliation{\TuB}
\author{Ria G. Krämer}
\affiliation{\ACP}
\author{Thomas K. Bracht}
\affiliation{\TUdo}
\author{Saimon F. Covre da Silva}
\affiliation{\JKU}
\affiliation{\brazil}
\author{Daniel Richter}%
\affiliation{\ACP}
\author{Stefan Nolte}%
\affiliation{\ACP}
\affiliation{\IOF}
\author{Armando Rastelli}%
\affiliation{\JKU}
\author{Doris E. Reiter}
\affiliation{\TUdo}
\author{Gregor Weihs}
\affiliation{\UIBK}
\author{Tobias Heindel}
\affiliation{\TuB}
\author{Vikas Remesh}
\affiliation{\UIBK}

\makeatletter
\patchcmd{\frontmatter@RRAP@format}{(}{}{}{}
\patchcmd{\frontmatter@RRAP@format}{)}{}{}{}
\renewcommand\Dated@name{}
\makeatother

\date{Date: \today \\ \phantom{XXX} E-mail: yk441@cam.ac.uk}

\begin{abstract}
The generation of single photons using solid-state quantum emitters is pivotal for advancing photonic quantum technologies, particularly in quantum communication. As the field continuously advances towards practical use cases and beyond shielded laboratory environments, specific demands are placed on the robustness of quantum light sources during operation. In this context, the robustness of the quantum light generation process against intrinsic and extrinsic effects is a major challenge. Here, we present a robust scheme for the coherent generation of indistinguishable single-photon states with very low photon number coherence (PNC) using a three-level system in a semiconductor quantum dot. Our novel approach combines the advantages of adiabatic rapid passage (ARP) and stimulated two-photon excitation (sTPE). We demonstrate robust quantum light generation while maintaining the prime quantum-optical quality of the emitted light state. Moreover, we highlight the immediate advantages for the implementation of various quantum cryptographic protocols.
\end{abstract}

\maketitle

\section*{Introduction}
Single photons used as flying qubits will play a vital role in the next generation of quantum technologies, enabling numerous applications from quantum communication to optical quantum computing \cite{kok2007linear, kimble2008quantum}. Semiconductor quantum dots (QDs) are a promising platform for quantum information processing due to their exceptional photon properties \cite{frick2023single,vajner_quantum_2022}. While photons can be generated with close-to-ideal properties using QDs in well-controlled laboratory environments, practical applications require a reliable operation also in realistic use-cases, posing a significant challenge to date.
To ensure robust operation in real-world settings, it is crucial that the QD photon source behaves robust against to fluctuations in both power and wavelength of the excitation laser.

Several optical excitation schemes have been developed to generate single-photon states from QDs \cite{luker2019review,Somaschi2016, bracht2021swing,karli2022super,koong2021coherent,wilbur2022notch, Thomas2021b, Vyvlecka2023, sbresny2022stimulated}. Here, the resonant two-photon excitation (TPE) of the biexciton  state stands out due to its potential for producing high-quality single photons with only spectral filtering required \cite{hanschke_quantum_2018,akopian_entangled_2006,schimpf_quantum_2021,Kappe_2023}.However, TPE has practical disadvantages, such as sensitivity to laser pulse parameters \cite{Kappe_2023, Ramachandran2024_multi}, limitations in indistinguishability \cite{scholl2020crux}, and a loss of control over photon number coherence (PNC). The latter is crucial for achieving security and reliability in various quantum communication protocols \cite{bozzio2022enhancing, karli2023controlling}.


In this work, we implement a robust excitation scheme for the generation of single-photon states from QDs with high indistinguishability and very low PNC. We begin by introducing our scheme, which combines the advantages of Adiabatic Rapid Passage (ARP) \cite{Kappe_2023,simon_robust_2011,wu_population_2011,debnath_high-fidelity_2013,kaldewey_coherent_2017,mathew2014subpicosecond} and stimulated-TPE (sTPE) \cite{akimov2006stimulated, sbresny2022stimulated,wei2022tailoring, karli2023controlling,yan2022double}, demonstrating robustness in excitation without compromising photon quality. This proposed excitation scheme, abbreviated as sARP (Stimulated ARP-enabled two-photon excitation), is then thoroughly analyzed and demonstrated experimentally. Finally, we discuss the advantages of sARP in different quantum communication scenarios.

\section*{Results and Discussion}
\begin{figure}[!h]
    \includegraphics[width=1\linewidth]{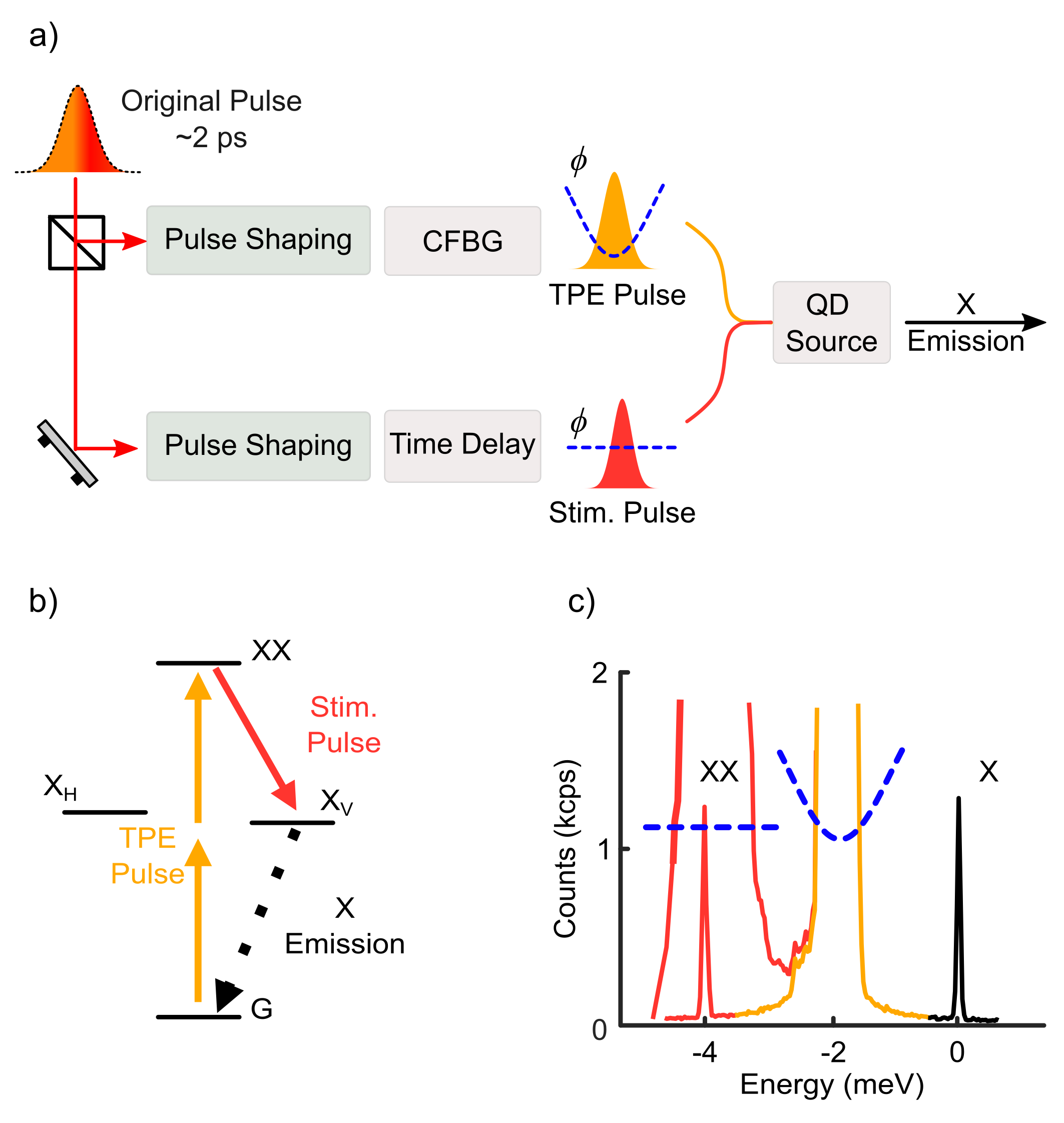}
    \caption{
    \textbf{(a)} Schematic of the sARP scheme. An initial laser pulse of \SI{2}{\pico\second} temporal width is shaped into two separate pulses resonant with the TPE and the $\mathrm{XX}$-$\mathrm{X}$ transition energy. The TPE pulse is chirped using a chirped fiber Bragg grating (CFBG), and both pulses are combined and directed onto a quantum dot (QD). The blue dashed lines represent the phase profile $\phi$ creating the chirp. The time delay between the pulses is precisely controlled via a delay stage. The collected emission, together with any stray light, is coupled to single-mode fibers and sent to a homemade monochromator for spectral filtering. \textbf{(b)} Energy level diagram illustrating the sARP. The biexciton state is initially prepared via ARP. In the absence of the stim. pulse, the biexciton state subsequently decays, emitting photons corresponding to the $\mathrm{XX}$ and $\mathrm{X}$ states through two possible decay channels. When a stimulation pulse is present, the $\mathrm{XX}$ is deterministically stimulated to one of the $\mathrm{X}$ states ($X_H$ or $X_V$), depending on the polarization of the stimulation pulse.  \textbf{(c)} Spectra of the $\mathrm{XX}$ and $\mathrm{X}$ emissions together with TPE (orange) and stim. (red) pulses.
    }    \label{Fig:Fig1}
\end{figure}

In TPE, two photons are absorbed simultaneously to prepare the biexciton state \cite{Brunner1994}. Upon relaxation, the biexciton-exciton cascade decays sequentially, releasing two photons with slightly different energies. These energies correspond to the biexciton to exciton ($\mathrm{XX}$) and exciton to groundstate ($\mathrm{X}$) transitions (see Fig \ref{Fig:Fig1}b), and are different due to the biexciton binding energy. This energy difference allow for spectral filtering of the excitation laser without the need for cross-polarization techniques, thereby simplifying the experimental setup. 

The nature of TPE requires us to address two critical issues to utilize it effectively in single-photon-based applications. The first issue is the resonant nature of TPE, which is highly sensitive to laser pulse parameters. Precise tuning of the wavelength and the power of the excitation laser is crucial for successful excitation \cite{Kappe_2023}. These strict requirements pose a barrier to achieving consistent and stable photon generation in practical applications where the laser pulse generation may have imperfections. The second issue is the limitation in the indistinguishability of the $\mathrm{X}$ photons due to the time jitter caused by the finite $\mathrm{XX}$ lifetime. This time jitter primarily arises from the stochastic nature of the $\mathrm{XX}$ decay process \cite{scholl2020crux}.

To address these two issues, firstly, we enable the ARP mechanism using chirped laser pulses \cite{Kappe2024ChirpedPM} to ensure robust and stable biexciton state generation. ARP relies on the sweep of the instantaneous laser frequency across the QD resonance. This sweeping action ensures a stable transition to the biexciton state, even in the presence of imperfections in the excitation parameters such as power and energy fluctuations \cite{wilbur2022notch,Kappe_2023,wei_deterministic_2014,RemeshCFBG}. Secondly, we eliminate the time jitter by using a second laser pulse to stimulate the transition of the $\mathrm{XX}$ to the $\mathrm{X}$ state. In Figure \ref{Fig:Fig1} we show the sketch of the proposed sARP scheme. Initially, the laser source with a pulse duration of \SI{2}{\pico\second} is tuned close to the TPE resonance energy. By using two pulse shapers as illustrated in Figure \ref{Fig:Fig1}a, we slice two pulses from this pulse, that are resonant with the TPE and the $\mathrm{XX}$-$\mathrm{X}$ transition energies, respectively, called the TPE and stim. pulse. The TPE pulse is chirped to \SI{45}{\pico\second\squared} using a chirped fiber Bragg grating \cite{RemeshCFBG}. The intensities of the TPE and stim. pulses are individually controlled via electronic variable optical attenuators (VOA, V800PA, Thorlabs) and the arrival time of the stim. pulse is precisely controlled via a fiber optic delay line (ODL-300, OZ Optics). The two beams are combined at a 10:90 beamsplitter near the optical window of a closed-cycle cryostat (base temperature \SI{1.5}{\kelvin}, ICEOxford) where the sample is mounted on a three-axis piezoelectric stage (ANPx101/ANPz102, Attocube systems AG). Our sample consists of GaAs/AlGaAs QDs grown by the Al-droplet etching method with the $\mathrm{X}$ emission centered around \SI{795}{nm} \cite{huber2017highly,da2021gaas}. Further details on the setup and sample structure have been described elsewhere \cite{Kappe_2023,karli2023controlling, RemeshCFBG}. The QD emission is collected via the same path as the excitation, and $\mathrm{X}$ and $\mathrm{XX}$ photons are filtered using  notch filters (BNF-805-OD3, Optigrate) and directed to single photon detectors or to a single-photon sensitive spectrometer (Acton SP-2750, Roper Scientific) equipped with a liquid Nitrogen cooled charge-coupled device camera (Spec10 CCD, Princeton Instruments). 

To characterize the photon emission, including single-photon purity, preparation fidelity, indistinguishability, and photon-number coherence (PNC), we analyze the collected $\mathrm{X}$ and $\mathrm{XX}$ photons with various experimental setups. 
Further details of photon characterization are discussed in the relevant sections of the paper.\\

\noindent\textbf{Robustness}\\
\begin{figure*}[!ht]
    \includegraphics[width=1\linewidth]{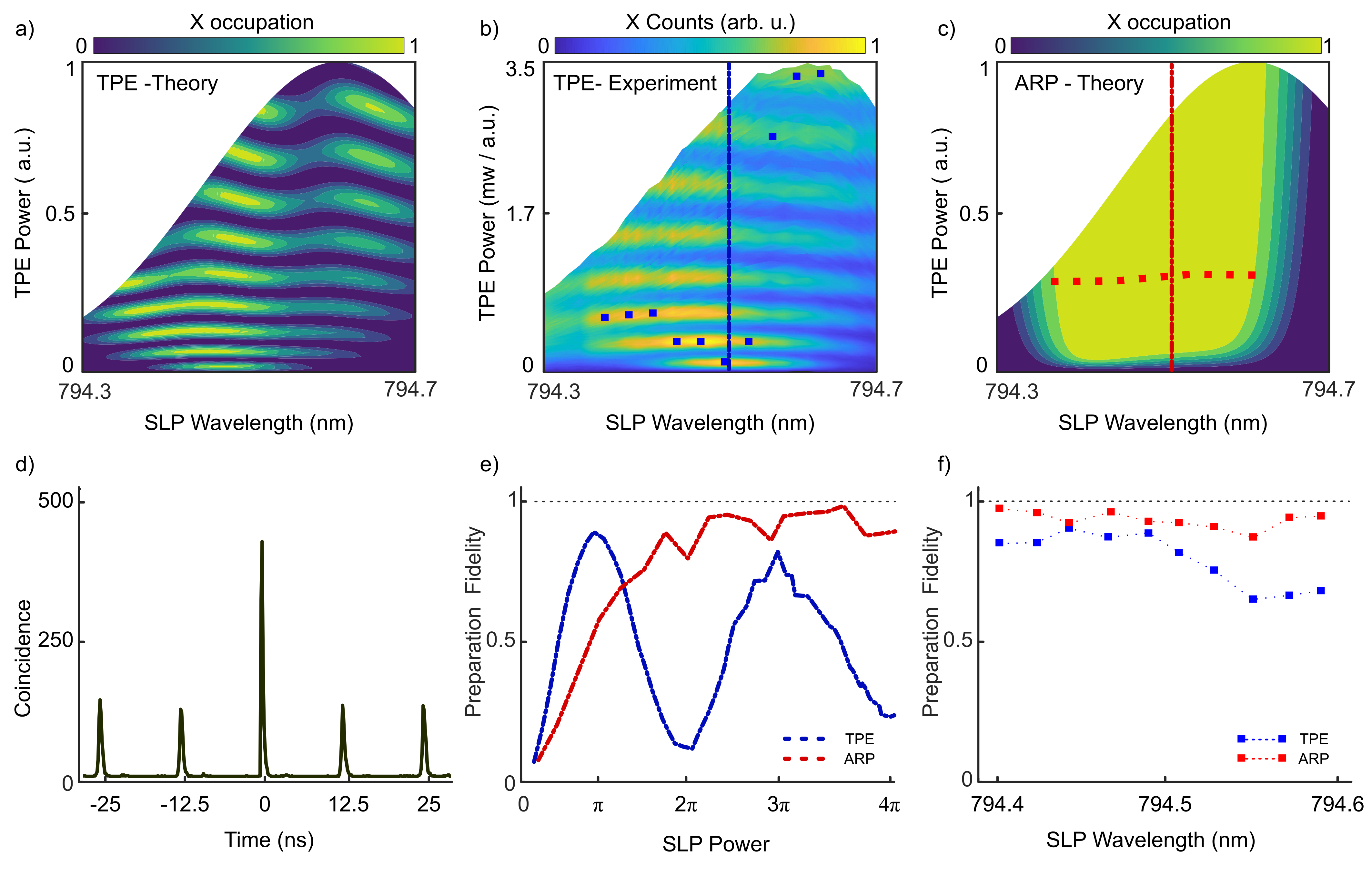}
    \caption{
    \textbf{(a)} Calculation of the $\mathrm{X}$ counts using a $\SI{6}{\pico\second}$ long shaped-laser-pulse (SLP) carved from a $\SI{2}{\pico\second}$ long initial Gaussian pulse. The x-axis shows the SLP center wavelength, the y-axis shows the SLP power, and the color represents the $\mathrm{X}$ counts. Considering the SLP is carved from an initial Gaussian pulse, the overall shape of the colormap exhibits a Gaussian profile as well.
    \textbf{(b)} Experimental results validating the calculated values presented in panel (a). Similar to the calculation, the experiment is conducted with an SLP derived from a $\SI{2}{\pico\second}$ long initial Gaussian pulse. The blue squares and blue dashed line indicate the parameters where preparation fidelities are calculated in panel (e) and (f).
    \textbf{(c)} Calculation of the $\mathrm{X}$ counts with an added chirp of ($\phi = \SI{45}{\pico\second\squared}$) to the SLP, demonstrating the effect of chirping on the $\mathrm{X}$ counts. The red squares and red dashed line represent the parameters selected to measure preparation fidelities in panel (e) and (f).
    \textbf{(d)} A representative cross-correlation measurement, with $\mathrm{XX}$ photon starting the clock and $\mathrm{X}$ photon stopping it.
    \textbf{(e)} Preparation fidelity measurement with different SLP powers for a fixed SLP wavelength, as indicated by the dashed lines in panels (b) and (c). The blue line represents the TPE scenario, where preparation fidelity is highly dependent on the laser pulse power, exhibiting peaks and troughs as the power varies.The red line represents the ARP scenario, showing an increase in preparation fidelity at low power, which then becomes flat and stable as the power increases, indicating robustness against to power fluctuations.
    \textbf{(f)} Preparation fidelity measurement for different SLP center wavelengths for the parameters shown as squares in panels (b) and (c). With TPE (blue squares) the QD experiences a drop in fidelity for red-detuned wavelengths, whereas ARP (red squares) maintains high preparation fidelity.
}
    \label{Fig:2Dmap}
\end{figure*}


Before introducing the stimulation pulse, we first discuss the importance of ARP to better illustrate the concept of robustness in the excitation process. Although the ARP mechanism has been widely studied and demonstrated to achieve robustness against power and wavelength imperfections, we would like to highlight another aspect where ARP offers advantages, further enhancing the overall robustness. Exciting a QD with TPE usually involves shaping spectrally narrow pulses from spectrally broad laser pulses. This is necessary because the biexciton binding energy of commonly used QDs is only a few \SI{}{\milli\electronvolt} (here \SI{4}{\milli\electronvolt}) corresponding to a wavelength difference between $\mathrm{X}$ and $\mathrm{XX}$ photons of only a few \SI{}{\nano\meter} (here about \SI{2}{\nano\meter}) and the excitation laser must be resonant with half of the biexciton state energy (see Figure \ref{Fig:Fig1}b for TPE level diagram and Figure \ref{Fig:Fig1}c for a representative spectrum). To achieve this, we shape approximately \SI{6}{\pico\second} long pulses from an initial \SI{2}{\pico\second} pulse, targeting the TPE energy. Using longer pulses for excitation also helps improve the spectral filtering of the emitted photons. However, using shaped laser pulses introduces a set of parameters that needs to be optimized for ideal excitation.
Figure \ref{Fig:2Dmap}a shows the calculated $\mathrm{X}$ counts under TPE. We performed calculations using a four-level model that includes the ground state, two exciton states with orthogonal linear polarizations, and the biexciton state. We solve the equations of motion numerically using a state-of-the-art numerical approach based on product tensor methods \cite{cygorek2022simulation,cygorek2024sublinear}. While the calculations shown in Figure \ref{Fig:2Dmap}(a) and (c) are performed without additional dephasing mechanisms, the inclusion of exciton-phonon interaction for longitudinal acoustic phonons is straight forward and becomes necessary when applying negative chirp to the pulse \cite{Kappe2024ChirpedPM, luker_influence_2012}. 
A linearly polarized laser pulse of Gaussian shape and approximate temporal duration of \SI{2}{\pico\second} full width at half maximum (FWHM) is spectrally shaped in a 4-f pulse-shaper. The effect of this shaping is incorporated into the simulation via a filter function, constructed by convolution of the pulse-shaper's spatial spread, assumed to be Gaussian, and the transmission function of a mechanical slit aperture. The resulting pulses are non-Gaussian and of approximately \SI{6}{\pico\second} temporal duration FWHM.    
For clarity, we refer to this as the shaped-laser-pulse (SLP).In Figure \ref{Fig:2Dmap}a, the SLP center wavelength is varied along the x-axis, and its power is varied along the y-axis, while the z-axis (represented by color) indicates simulated $\mathrm{X}$ counts. For details of the model used in the calculation, see \cite{Kappe_2023}. In Figure \ref{Fig:2Dmap}b, we present the measured data obtained using the same initial pulse and pulse shaping parameters. Our results indicate that, to achieve maximum $\mathrm{X}$ counts, both the energy corresponding to the center wavelength of SLP and the initial unshaped pulse must be resonant with the TPE energy. Otherwise, the maximum in $\mathrm{X}$ counts shifts to different SLP wavelength and power values, leading to unexpected Rabi-rotation behavior\cite{Ramsay2007_square}. However, employing ARP offers a solution in this context. In Figure \ref{Fig:2Dmap}c, we calculate the same $\mathrm{X}$ counts as in Figure \ref{Fig:2Dmap}a but with an additional \SI{45}{\pico\second\squared} chirp on the SLP. The colormap shows a wide range of SLP wavelengths and powers where the counts peak. The red dots marked in Figure \ref{Fig:2Dmap}c are measured experimentally to demonstrate that the $\mathrm{X}$ counts reach the maximum.    

Another crucial aspect to consider is the preparation fidelity, which refers to the likelihood of the QD being excited to the biexciton state through TPE or ARP. To evaluate this, we conduct preparation fidelity measurements using a cross-correlation coincidence experiment, focusing on the cascaded emission of the $\mathrm{XX}$ and the $\mathrm{X}$ photos\cite{Vajner2023OnDemandGO, wang_-demand_2019}. In our experimental setup, $\mathrm{XX}$ and $\mathrm{X}$ photons are spectrally filtered, coupled into single-mode fibers and directed towards a superconductor nanowire single photon detector (SNSPD, Eos, Single Quantum). The arrival times of these photons are recorded using a time tagger(time-tagger Ultra, Swabian Instruments), and we employ coincidence counting to ascertain the correlation between the emissions. Figure \ref{Fig:2Dmap}d presents a representative cross-correlation histogram from the measurements. The cross-correlation histogram depicts the arrival time differences between the $\mathrm{XX}$ and the $\mathrm{X}$ photon. The clock starts with the detection of the $\mathrm{XX}$ photon and stops with the detection of the $\mathrm{X}$ photon. Then, the preparation fidelity ($\mathcal{F}_p$) of the biexciton state is extracted using the following formula
\begin{equation}
\mathcal{F}_p=\frac{\mathrm{A_{side}}}{\mathrm{A_{center}}} \times \mathrm{C_{pol}}.    
\end{equation}
where $\mathrm{A_{center}}$ corresponds to the area of the center (zero delay) peak, and $\mathrm{A_{side}}$ corresponds to the mean area of the uncorrelated peaks and $\mathrm{C_{pol}}$ is the correction factor due the polarization-selective collection. In our case, as the emission is circularly polarized, the correction factor is 2 \cite{Vajner2023OnDemandGO, wang_-demand_2019}.

In Figure \ref{Fig:2Dmap}e, we show the computed $\mathcal{F}_p$ values for different SLP powers for a fixed SLP wavelengths shown as dashed lines in Figure \ref{Fig:2Dmap}b (blue line) and Figure \ref{Fig:2Dmap}c (red line) respectively for TPE and ARP. In TPE, a high $\mathcal{F}_p$ value is measured when the SLP power is set to $\pi$, whereas ARP reaches a plateau starting from around $2\pi$ onward.
We also measured $\mathcal{F}_p$ values for different SLP wavelengths in panel f. The parameters for TPE are marked in Figure \ref{Fig:2Dmap}b with blue squares, and the parameters for ARP are marked in Figure \ref{Fig:2Dmap}c with red squares. Unlike in the ARP case, we selected different power values for TPE (blue squares) to measure $\mathcal{F}_p$, where the $\mathrm{X}$ counts reach the maximum at each SLP wavelength. In TPE, we observe a variation in $\mathcal{F}_p$ resulting from changes in the SLP wavelength. Especially as the laser becomes red-detuned, there is a noticeable decrease in preparation fidelity. In contrast, ARP (red squares) maintains a higher preparation fidelity within the varied wavelengths of SLP.


\noindent\textbf{sARP Scheme}\\
Although ARP provides robust biexciton state preparation, the indistinguishability of $\mathrm{X}$ photons remains limited due to the time jitter caused by the finite $\mathrm{XX}$ lifetime. In addition, control over the PNC is also lost due to the incoherent decay process of the $\mathrm{XX}$ \cite{karli2023controlling}. To address this, we introduce a second laser pulse that arrives approximately \SI{7}{\pico\second} after the TPE pulse and is resonant with the $\mathrm{XX}$-to-$\mathrm{X}$ transition energy, which stimulates the $\mathrm{XX}$ to the $\mathrm{X}$ state \cite{akimov2006stimulated}. Successful stimulation of the $\mathrm{XX}$ requires the stim. pulse parameters to be optimized for arrival time, power, and polarization. Further details on stim. pulse optimization can be found in \cite{karli2023controlling}. When the $\mathrm{XX}$ decay is successfully stimulated, this reduces the time jitter and results in high indistinguishability \cite{sbresny2022stimulated,yan2022double,wei2022tailoring,karli2023controlling} as well as the ability to tailor the amount of PNC for the $\mathrm{X}$ photons \cite{karli2023controlling}.

\begin{figure}[H]\includegraphics[width=1\linewidth]{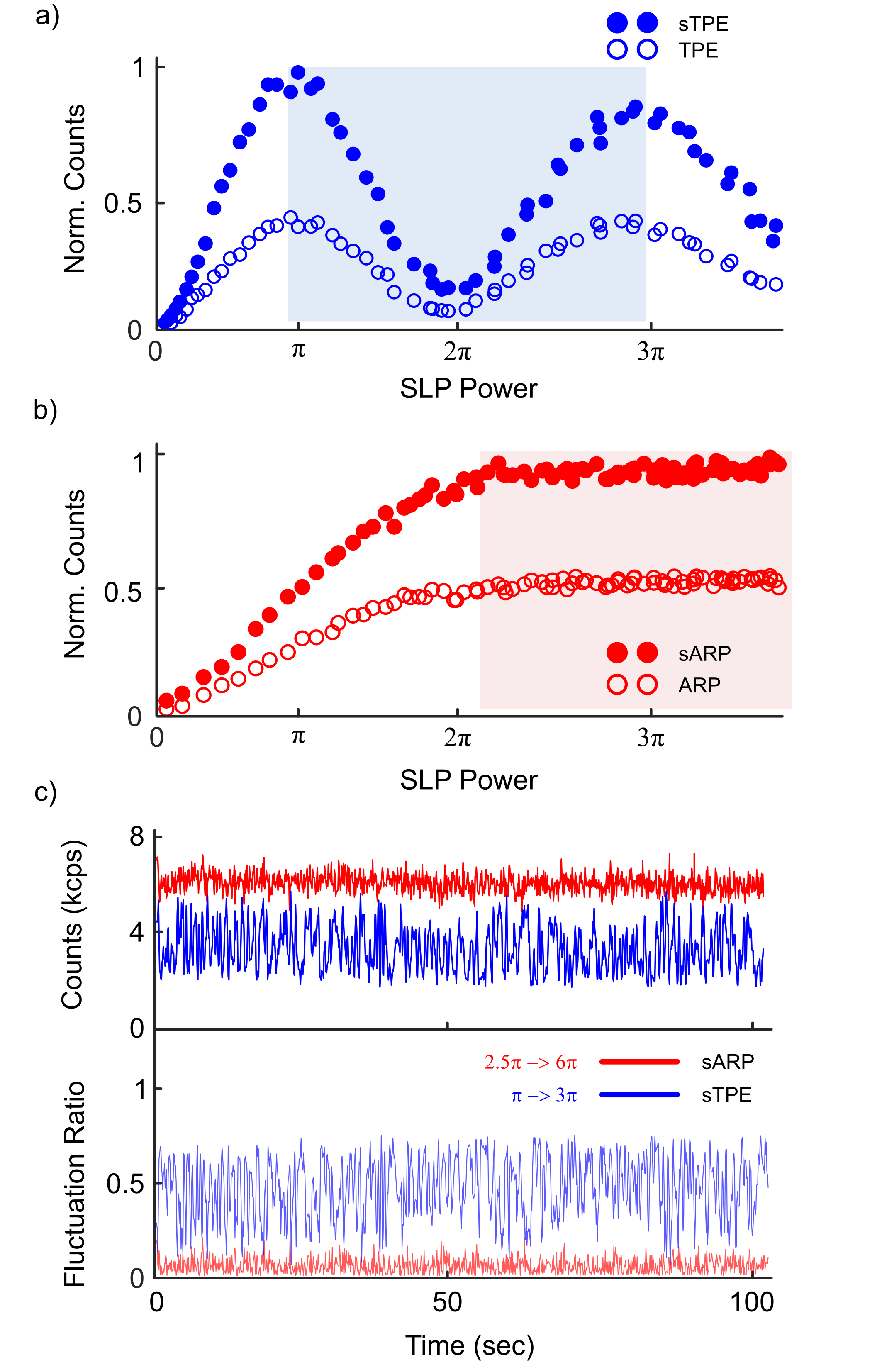}
\caption{\textbf{(a)} Rabi rotations of the $\mathrm{X}$ photons for TPE (blue circles) and sTPE (blue filled circles) without chirp. The counts are enhanced due to addition of the stim. pulse with the same polarization as the collection polarization basis. The blue square indicates the range of power fluctuations discussed in panel (c). 
\textbf{(b)} $\mathrm{X}$ counts reach a plateau with chirped SLP at $2\pi$ pulse power. The red circles represent the ARP case, while the red filled circles represent the sARP case, both showing increased counts when $\mathrm{XX}$ is stimulated to the same polarization as the collection. The red square indicates the range of power fluctuations discussed in panel (c).   
\textbf{(c)} A 100 seconds window of recorded counts, each with a 100 ms integration time. The blue line in the top panel (sTPE) shows the sensitivity of the emission to excitation power fluctuations, with counts varying significantly as the pulse power is randomly swept from $\pi$ to $3\pi$ (indicated by the blue square in panel (a)). Conversely, the red line (sARP) indicates a stable count rate, remaining unaffected by pulse power variations from $2.5\pi$ to $6\pi$ (indicated by the red square in panel (b)). The bottom panel shows the corresponding absolute value of the relative deviations from the mean for each respective point in the top panel.
}
    \label{Fig:sTPE-rabi}
\end{figure}

Figure \ref{Fig:sTPE-rabi}a shows the Rabi rotations with no chirp ($\phi = \SI{0}{\pico\second\squared}$) the empty circles and filled circles represent the $\mathrm{X}$ counts recorded under TPE and sTPE respectively.In this scenario, both traces follow a similar pattern, but there is an increase in $\mathrm{X}$ counts under sTPE due to the stimulation of the $\mathrm{XX}$ decay into the collected polarization basis \cite{karli2023controlling}.
Figure \ref{Fig:sTPE-rabi}b shows the $\mathrm{X}$ counts as in (a), with the only difference being that \SI{45}{\pico\second\squared} chirp was added to the SLP. We chose this value as the required chirp to enable ARP for our pulse parameters (see \cite{Kappe_2023} for details). When ARP is enabled, the $\mathrm{X}$ counts increases and then reaches a plateau after $2\pi$ SLP power. Upon $\mathrm{XX}$ stimulation (filled circles), we observe an increase in the $\mathrm{X}$ counts, proving that stimulation is successful together with the ARP. 

Additionally, in Figure \ref{Fig:sTPE-rabi}c, we show the robustness of sARP in practical scenarios to demonstrate robust single-photon emission. The top panel of Figure \ref{Fig:sTPE-rabi}c shows the recorded count rate for 100 seconds by an SNSPD with a 100 ms integration time where the SLP power is varied randomly from $\pi$ to $3\pi$ for sTPE (blue line) and ($2.5\pi$) to  ($6\pi)$ for sARP(red line).
The randomization is achieved by applying random voltages to an electronic variable optical attenuator (eVOA, V800PA - Thorlabs). The bottom panel of Figure \ref{Fig:sTPE-rabi}c shows the absolute value of the relative deviations from the mean for each respective point in the top panel. sARP exhibits a robust and stable count rate as a function of time when the pulse power is varied. Despite this large variation, the detected counts remain remarkably constant, exhibiting an average deviation of $5(1)\%$ from the mean photon counts of the plateau, compared to a $44(2)\%$ average deviation for sTPE. \\

\noindent\textbf{Photon Characterization}\\
To certify the generated single-photon quality, we employ a Hanbury-Brown and Twiss (HBT) setup to perform $g^{(2)}(\tau)$ measurements and a Hong-Ou-Mandel (HOM) setup to measure indistinguishability at the $2.5\pi$ SLP power. The obtained $g^{(2)}(0)$ values are $0.0081(3)$ and $0.0052(2)$ respectively for ARP and sARP, and indistinguishabilities of $V_{\text{ARP}} = 0.55(1)$ and $V_{\text{sARP}} = 0.80(1)$. For details of HBT and HOM experiments, see Supplementary Information.

\begin{figure}[!h]
    \includegraphics{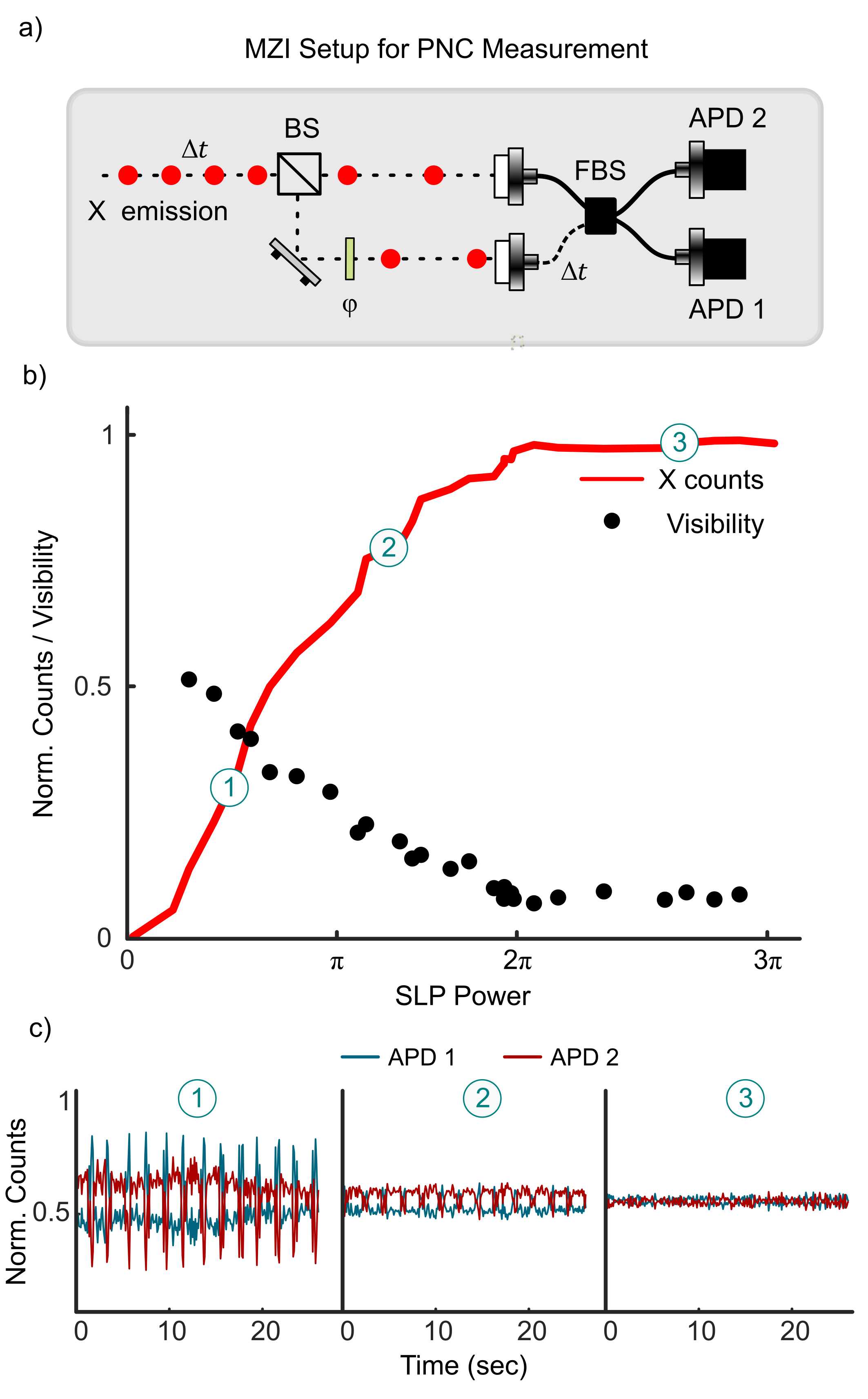}
\caption{\textbf{(a)} Schematic of the experimental setup for measuring photon-number coherence (PNC) using a Mach-Zehnder interferometer (MZI). Collected $\mathrm{X}$ photons is directed through a 50:50 beamsplitter (BS), creating two paths with a time difference $\Delta t$ between photons. One path includes a phase scan apparatus ($\phi$) to control and scan the phase. The paths are then recombined at a fiber beam splitter (FBS), and the outputs are detected by two avalanche photodiodes (APD1 and APD2). The setup allows for the measurement of visibilities between the counts at APD1 and APD2 as the phase is scanned.
\textbf{(b)} The red line shows the $\mathrm{X}$ counts as the SLP power is scanned, while the black dots represent the visibility derived from PNC measurements. The visibility drops as the SLP power increases and saturates when the counts reaches the plateau.
\textbf{(c)} Exemplary time-resolved photon count rates for three different SLP powers, marked by numbers 1, 2, and 3 in panel (b).
}
    \label{Fig:PNC}
\end{figure}

To calculate PNC, we record the photon counts in the output ports of a phase-evolving MZI (see Figure \ref{Fig:PNC}a for a sketch of setup) and extract the visibilities as described in \cite{loredo2019generation,karli2023controlling}. Figure \ref{Fig:PNC}b shows the extracted visibilities as black dots and the corresponding collected $\mathrm{X}$ counts as a red line. In panel c, the recorded counts at the fiber beam splitter outputs are displayed as a function of time, for the SLP power levels marked with numbers 1, 2, and 3 in panel b. We observe that the extracted visibilities (black dots) start at high values for low SLP powers and reach a minimum value close to zero, where the photon counts reach the maximum. Afterwards, any power fluctuations do not affect the recorded visibility, which remains near zero and proportional to the PNC when preparation fidelity is high. This implies that as the excitation process enables the ARP, both the single-photon counts remain stable, and the PNC is absent.

\subsection*{Advantages of sARP for Quantum Information}
As demonstrated above, the sARP scheme enables us to combine excellent quantum-optical properties of the generated photons with robustness against environmental fluctuations, which is highly beneficial in real-world applications. In the following we discuss the advantages that this robustness offers for the implementation of different cryptographic protocols.\\

\noindent\textbf{Enhanced Security in Quantum Key Distribution}\\
Quantum Key Distribution (QKD) aims to establish a secure key between two remote parties sharing a quantum channel and an authenticated public classical channel \cite{gisin2002quantum,gisin2007quantum,pirandola2020advances}. While QKD protocols can be provably secure, in an information-theoretical sense, practical implementations often require certain assumptions for the devices used. Despite the existence of single-photon-based QKD implementations \cite{vajner_quantum_2022} and progress towards fully device-independent QKD, practical, assumption-free secure schemes remain beyond current technology \cite{zapatero2023advances}. Experimental assumptions can be exploited by eavesdroppers to leak information through side channels, enabling quantum hacking strategies \cite{jain2016attacks}.

\begin{figure*}[!ht]
\centering
    \includegraphics[width=\textwidth]{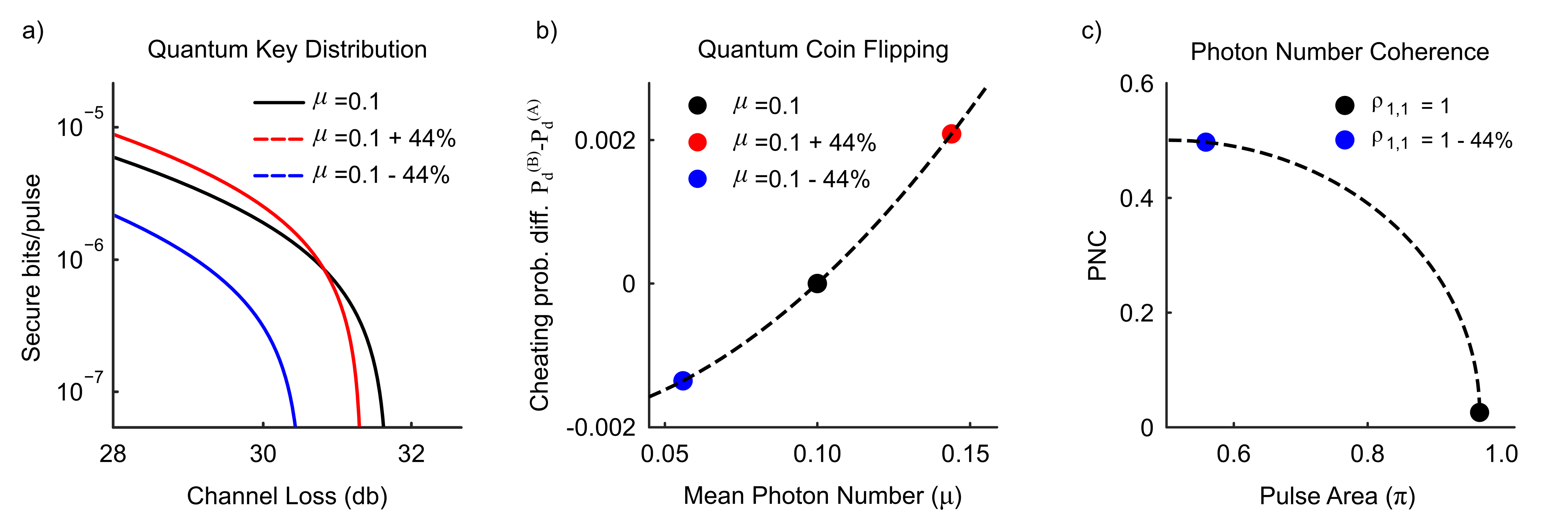}
    \caption{Examples of the advantage of robust fluctuation-free excitation in different quantum communication scenarios.  \textbf{(a)} Influence of power-fluctuations on secure key rate in BB84-QKD, calculated with parameters from Table \ref{tab:parameters} in finite-size regime. Higher (red line) or lower (blue line) $\mu$ than assumed by the communicating parties, due to power fluctuations, might render the key insecure. \textbf{(b)} Influence of power-fluctuations on the fairness of the strong coin flipping protocol as the difference in cheating probabilities ($P_d^{(B)}-P_d^{(A)}$) increases in the presence of fluctuations in $\mu$ \cite{pappa2011practical}. \textbf{(c)} For coherent excitation, deviations from the ideal $\pi$-pulse due to power fluctuations result in considerable PNC, reducing the security of cryptographic protocols like BB84 and coin flipping \cite{bozzio2022enhancing}.
    }
    \label{fig:QKD}
\end{figure*}

In standard BB84-type QKD schemes \cite{bennet1984proceedings} for instance, it is assumed that the average photon number per pulse, $\mu$, sent from Alice to Bob, is constant and known. Uncontrolled fluctuations in the emission rate, as observed in the TPE case in Figure \ref{Fig:sTPE-rabi}c, however, will result in changes of the multi-photon probability and can therefore be exploited by an eavesdropper via photon-number-splitting attacks (PNSA)\cite{lutkenhaus2002quantum}. QKD security proofs typically compensate for PNSA using the tagging model, adjusting the amount of necessary privacy amplification \cite{gottesman2004security}. In this case the secure key rate is determined by a version of the GLLP equation \cite{gottesman2004security}. The multi-photon probability per pulse, $\mathrm{P_m}$, can be bounded by $\mathrm{P_m} \leq \frac{1}{2} \mu^2 g^{(2)}(0)$ \cite{waks2002security}, i.e., any signal fluctuation leads to quadratic changes in multi-photon probability. Figure \ref{fig:QKD}a illustrates the calculated expected secure bit fraction $r$ in the standard BB84 protocol using parameters from Table \ref{tab:parameters} and calculated in the finite-size regime \cite{cai2009finite}.
Although the intrinsic property of preparation fidelity is nearly unity, we choose $\mu = 0.1$ (the solid black line in Figure \ref{fig:QKD}), which is a typical value in realistic implementations. This choice considers parameters such as extraction efficiency, collection efficiency, and losses in the optical components\cite{munzberg2022fast}. Even in cases where record end-to-end efficiencies of single-photon emitters exceed 0.5 \cite{tomm_bright_2021}, the efficiency would still be reduced due to losses during state preparation on the sender side.
Without robust excitation, power fluctuations cause the preparation fidelity to fluctuate (cf. Figure \ref{Fig:2Dmap}e), inducing changes in $\mu$ by $\pm$44~\%. 
Furthermore, we choose $R$ and $g^{(2)}(0)$ based on the experimental parameters above. The accumulation time matches satellite fly-by times and reasonable post-processing block sizes. The protocol parameters $f$, $ q$, $Q_{\text{key}}$ are standard values for a symmetric BB84 QKD protocol without pre-attenuation, while the values for the security parameters and experimental assumptions on the detector dark counts and the detection module transmission are typical values for QKD implementations \cite{gao2023atomically,morrison2023single}.
\\
The red and blue lines in Figure \ref{fig:QKD}  represent fluctuations of $\mu \pm 44\%$, which is the average deviation of TPE under randomly induced power fluctuations, as shown in Figure \ref{Fig:sTPE-rabi}c. 
A different source of fluctuation could be an eavesdropper attempting to alter the excitation power by injecting a laser from the outside. 
An increase in $\mu$ increases the rate at low loss, but reduces the tolerable loss due to more multi-photon events that can be used for photon number splitting attacks, while a decrease in $\mu$ decreases the amount of secure key that is exchanged. As a consequence, random power fluctuations can render keys insecure if the parties operate in the high-loss regime and are unaware of these fluctuations.
In QKD, conservative assumptions are mandatory. An eavesdropper could block pulses during periods of lower $\mu$, only allowing those with higher $\mu$ to pass, thereby increasing the multi-photon probability, rendering the protocol insecure. When Alice and Bob operate at the $\pi$-pulse condition, fluctuations only decrease $\mu$. However, $\mu_{\text{meas}}$ measured at an expected $\pi$-pulse might already be affected by power fluctuations, making it smaller than $\mu_{\text{real}}$. An eavesdropper could exploit this by selectively transmitting pulses after non-demolition measurements of the mean-photon-number, effectively increasing $\mu$.
Active monitoring of mean-photon-number could mitigate fluctuations in this parameter, but this adds experimental complexity, particularly in multi-party protocols with uncorrelated random fluctuations. Decoy states, standard in WCP QKD but also applicable to single-photon implementations \cite{bozzio2022enhancing}, become ineffective with intrinsic fluctuations of $\mu$. Hence, our chirped excitation scheme, which ensures stable emission rates even under power fluctuations, enhances the security of future QKD implementations.

\begin{table}[]
\caption{Parameters used for calculating Figure \ref{fig:QKD}}
\label{tab:parameters}
\begin{tabular}{l|l}
Parameter  & Value  \\ \hline
Mean photon number $\mu$ & 0.1 \\
Laser repetition rate $R$  & $80\,$MHz    \\
Accumulation time $T$   & $100\,$s     \\
Error correction code efficiency $f$  & $1.2$         \\
Constant sifting factor $q$  & $0.5$         \\
Detection module error $e_{\text{det}}$ & $0.02$     \\
Dark count probability $p_{dc}$ & $10^{-7}$     \\
Security parameters $\epsilon_{EC}= \epsilon_{PA}= \epsilon_{PE}= \tilde{\epsilon}$ & $10^{-9}$     \\
$g^{(2)}(0)$ & $0.005$ \\
Fraction of bits used for key $Q_{\text{key}}$ & $0.9$   \\ \hline
Coin flipping protocol parameter $a$ & $0.9$  \\
Coin flipping protocol rounds $K$ & $500$  \\ 
\end{tabular}
\end{table}

\noindent\textbf{Improved Fairness in Quantum Coin Flipping}\\
Moreover, also cryptographic primitives beyond QKD are affected by signal fluctuations in the quantum channel. For example, in the strong quantum coin flipping protocol that generates a mutually unbiased random bit between two parties in a distrustful setting \cite{pappa2011practical}, fluctuations of $\mu \pm 44\%$ impair the protocols performance. While Bob’s cheating probability $P_d^{(B)}$ depends on the multi-photon probability $P_m (\mu)$, Alice's cheating probability $P_d^{(A)}$ remains constant when $\mu$ changes, which allows Bob to cheat by observing signal fluctuations and only allowing pulses with high $\mu$ that yield more multi-photon events. Figure \ref{fig:QKD}b depicts calculations of the difference between Alice's and Bob's cheating probabilities as a function of $\mu$. This results in an unfair advantage of Bob for higher $\mu$ and an advantage for Alice for smaller $\mu$, with even a small difference leading to significant consequences in applications that rely on the fairness of the coin flip. Here, we have chosen the number of exchanged pulses per coin flip $K$ to ensure a quantum advantage and the state preparation parameter $a$ to ensure initial fairness without power fluctuations.
\\

\noindent\textbf{Constant amount of PNC}\\
Most quantum cryptography protocols require the absence of PNC to avoid side-channel attacks \cite{bozzio2022enhancing}. The stimulated two-photon excitation in principle yields PNC, unless a perfect $\pi$-pulse is used. Therefore, any fluctuation in excitation power that alters the perfect $\pi$-pulse condition will lead to PNC, as shown in Figure \ref{fig:QKD}c, and by that reduce the security of the quantum cryptography protocol. But using the robust excitation shown in this work maintains the high preparation fidelity and ensures the absence of PNC even in the presence of power fluctuations (cf. Figure \ref{Fig:PNC}), guaranteeing security.

\subsection*{Conclusions}
In conclusion, this study advances QD-based single-photon generation by combining the high photon indistinguishability achieved via sTPE with the robust state preparation of ARP. We demonstrated that robust biexciton state preparation with highly indistinguishable $\mathrm{X}$ photons are vital for maintaining optimal performance and security in quantum communication protocols, especially in scenarios involving fluctuating emission rates and PNC.

Moreover, our findings pave the way for compact and robust quantum photonic systems with potential for practical device integration, such as fiber-coupled single-photon sources \cite{Rickert_2024_pigtail,Northeast2021_fibernanowire}, by leveraging recent developments in CFBGs \cite{RemeshCFBG} and laser miniaturization technologies, including integrated titanium
lasers \cite{Yang2024_titan_laser} and high-repetition-rate mode-locked semiconductor lasers \cite{Schlehahn2015, Mangold2014}. In combination with compact cryocoolers for user-friendly low-temperature operation of the quantum emitters\cite{Gao2022}, these technologies show prospects for the field-deployment of advanced and robust quantum light sources beyond shielded laboratory environments. Our work thus lays a solid foundation for the integration and miniaturization of scalable quantum devices, bringing practical quantum technologies closer to widespread real-world applications.

\section*{Acknowledgments}
YK, RS, FK, VR, and GW acknowledge the financial support through the Austrian Science Fund (FWF) projects with Grant DOIs 10.55776/TAI556 (DarkEneT), 10.55776/W1259 (DK-ALM Atoms, Light, and Molecules), 10.55776/FG5, 10.55776/I4380 (AEQuDot), 10.55776/COE1 (quantA), and the infrastructure funding from FFG (HuSQI). D.A.V. and T.H. acknowledge financial support by the German Federal Ministry of Education and Research (BMBF) via the project “QuSecure” (Grant No. 13N14876) within the funding program Photonic Research Germany, and the BMBF joint project “tubLAN Q.0” (Grant No. 16KISQ087K). TKB and DER acknowledge financial support from the German Research Foundation DFG through project 428026575 (AEQuDot). RGK, DR and SN acknowledge financial support from the German Federal Ministry of Education and Research through project 13N16028 (MHLASQU) and the German Research Foundation DFG (455425131, OH-SUPER and 448663633, fs2CVBG). A.R. and SFCdS acknowledge the FWF projects FG 5, P 30459, I 4320, the Linz Institute of Technology (LIT) and the European Union's Horizon 2020 research, and innovation program under Grant Agreement Nos. 899814 (Qurope), 871130 (ASCENT+) and the QuantERA II Programme (project QD-E-QKD). For open access purposes, the author has applied a CC BY public copyright license to any author accepted manuscript version arising from this submission.

\newpage
\appendix
\clearpage
\section{Single Photon Characterization}
\begin{figure}[t]\includegraphics{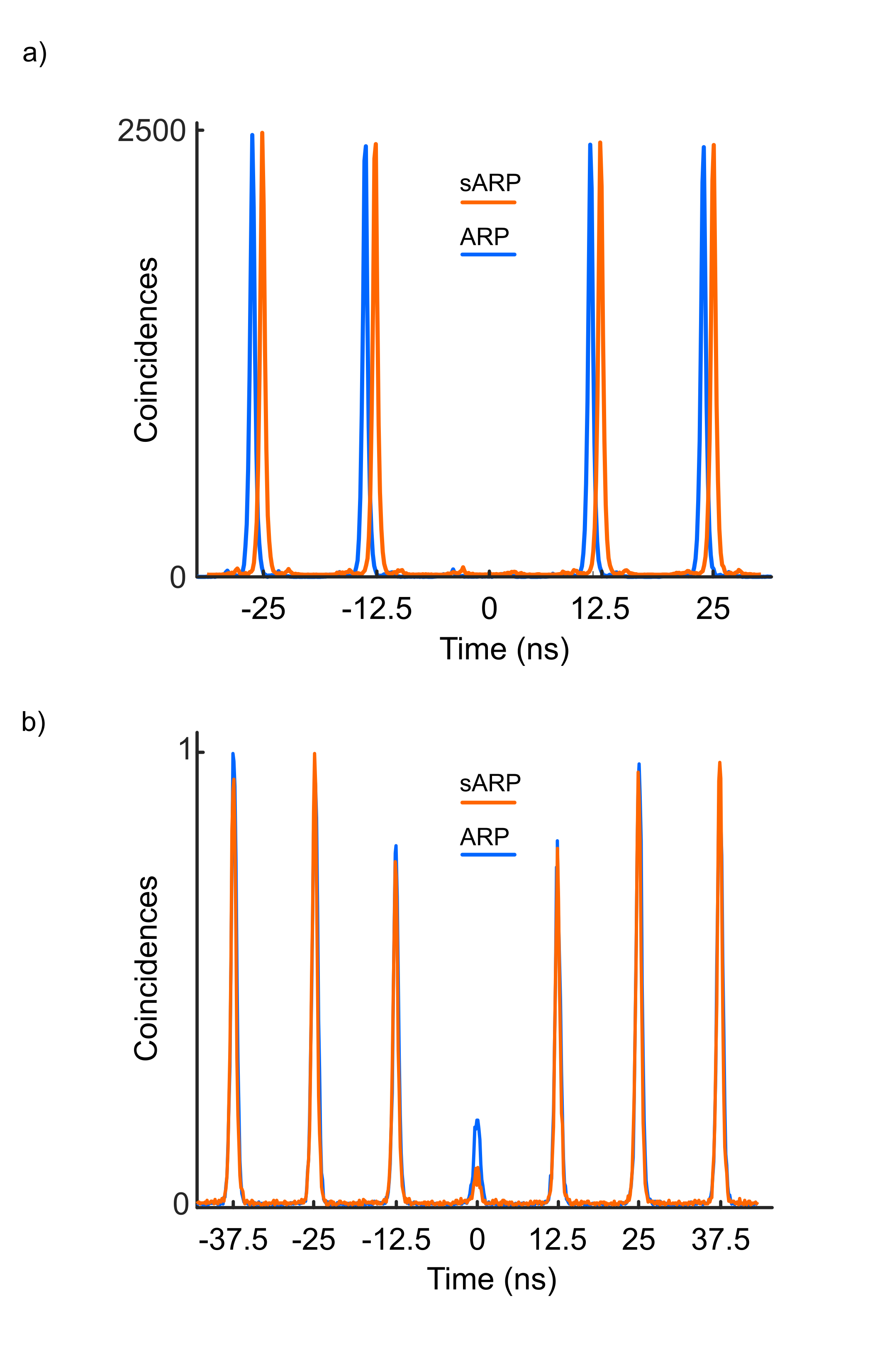}
\caption{\textbf{(a)} The second-order auto-correlation results of $\mathrm{X}$ photons show $g^{(2)}(0)$ values of $0.0081(3)$ and $0.0052(2)$ for ARP (blue line) and sARP (orange line), respectively. The lines are shifted slightly for better visualization. \textbf{(b)} HOM visibility measurements to calculate the indistinguishability of the emitted photons, $V_{\text{ARP}} = 0.55(1)$ and $V_{\text{sARP}} = 0.80(2)$. }
    \label{Fig:g2-hom}
\end{figure}

The $g^{(2)}(\tau)$ was measured for ARP and sARP when SLP power was set to \(2.5\pi\). In Figure \ref{Fig:g2-hom}a, our HBT measurement yields a $g^{(2)}(0)$ value of 0.0081(3) for ARP (blue line) and 0.0052(2) for sARP (orange line), thereby confirming the single-photon emission.

To quantify the indistinguishability, we conducted Hong-Ou-Mandel (HOM) measurements by directing the collected photons through a Mach-Zehnder Interferometer (MZI). This setup matches the time difference between its arms to the repetition rate of the excitation laser (80 MHz, corresponding to a 12.5 ns separation of pulses). Then, sequentially emitted photons interfere at a fiber beam-splitter. The outputs of the MZI are connected to two single-photon detectors, where we register the coincidences.

Figure \ref{Fig:g2-hom}b shows the HOM results (blue line for ARP, orange line for sARP). To obtain the HOM visibility as a measure of photon indistinguishability, we integrated the time window (\SI{2}{\nano\second}) for all the peaks in the histogram and calculated the area of the peaks.

Knowing that the center peak corresponds to the zero time delay, we have calculated the HOM visibility with the formula \cite{ollivier2020_HOM_side_peaks}
\begin{equation}
    V_{\text{HOM}} = 1 - 2(\frac{\mathrm{A_{\text{center}}}}{\mathrm{A_{\text{avg. of uncorr.}}}}).
\end{equation}
Where $\mathrm{A_{\text{avg. of uncorr.}}}$ is the average area of the uncorrelated peaks at $\pm$\SI{25}{\nano\second} and $\pm$\SI{37.5}{\nano\second}. The HOM visibilities are found to be $V_{\text{ARP}} = 0.55(1)$ and $V_{\text{sARP}} = 0.80(2)$. 

\newpage
\bibliography{Preprint_v3}

\end{document}